\documentclass[10pt,superscriptaddress,floatfix,twocolumn,amsfonts,amsmath,amssymb,nofootinbib,longbibliography,aps,prb]{revtex4-2}

\usepackage{graphicx} 
\usepackage{dcolumn} 
\usepackage{bm} 
\usepackage{float}
\usepackage{color}

\usepackage[colorlinks, breaklinks=true,linkcolor=red, citecolor=blue, linktocpage=true]{hyperref}

\begin{document}
	
	\title{Driving Viscous Hydrodynamics in Bulk Electron Flow in Graphene Using Micromagnets}

	\author{Jack N. Engdahl}
	\thanks{j.engdahl@student.unsw.edu.au}
	\affiliation{School of Physics, University of New South Wales, Sydney 2052, Australia}
		\author{Ayd\i n Cem Keser}
	\affiliation{CSIRO, Bradfield Road, West Lindfield NSW 2070, Australia}
\author{Thomas Schmidt }
\affiliation{Department of Physics and Materials Science, University of Luxembourg, Luxembourg}
\affiliation{School of Chemical and Physical Sciences, Victoria University of Wellington,\\ P.O. Box 600, Wellington 6140, New Zealand}
	\author{Oleg P. Sushkov}
	\affiliation{School of Physics, University of New South Wales, Sydney 2052, Australia}
	
	\date{\today}
	\begin{abstract}
		{We consider the hydrodynamic flow of an electron fluid in a channel formed in a two-dimensional electron gas (2DEG) with no-slip boundary conditions.
To generate vorticity in the fluid the flow is influenced by an array of micromagnets on the top of 2DEG.
We analyse the viscous boundary layer and demonstrate
anti-Poiseuille behaviour in this region. Furthermore we predict a  longitudinal voltage modulation, where a periodic magnetic field generates a voltage term periodic in the direction of transport. 
From the experimental point of view we propose a method for a boundary-independent measurement of the viscosity of different
electron fluids. The results are applicable to graphene away from the charge neutrality point and to
semiconductors.}
  
\end{abstract}

\maketitle
\section{Introduction}

It has been over half a century since the theory of viscous electron hydrodynamics was first developed \cite{gurzhi_hydrodynamic_1968} and in recent years viscous effects have been observed in the electron fluid experimentally~\cite{polini_viscous_2020, narozhny_hydrodynamic_2022}.
	Electron transport transitions to the hydrodynamic regime when the momentum-conserving electron-electron scattering length, $l_{ee}$, is the shortest length scale in the system and becomes much shorter than the momentum-nonconserving diffusive scattering length, $l_{\rm mfp}$, and the width of the sample, $W$. In this regime electron-electron interactions dominate and a local thermodynamic equilibrium is established among the electrons, enabling them to flow as a fluid.
Since this condition requires samples of very high purity, 
the theoretical prediction of the hydrodynamic regime in the 1960s~\cite{gurzhi_hydrodynamic_1968,gurzhi_minimum_1963} could only be realized experimentally after the development of high-mobility semiconductors~\cite{de_jong_hydrodynamic_1995} and graphene~\cite{lucas_hydrodynamics_2018,polini_viscous_2020}. 

Hydrodynamic electron flow manifests itself in a range of effects that have been observed.
While hydrodynamic electron transport was suggested to have been achieved already in 1995 in Ref.~\cite{de_jong_hydrodynamic_1995}, this result was immediately challenged by Gurzhi in Ref.~\cite{gurzhi_electron-electron_1995}: the conditions for hydrodynamics in the bulk material were satisfied, but they were not satisfied for hydrodynamics in a 2DEG. It was not until 2016 that reliable evidence of hydrodynamic transport was presented in Refs.~\cite{bandurin_negative_2016, moll_evidence_2016, alekseev_negative_2016}. Signatures of hydrodynamic behavior include the decrease of resistance with increasing temperature called the Gurzhi effect \cite{de_jong_hydrodynamic_1995,krishna_kumar_superballistic_2017,bandurin_fluidity_2018,keser_geometric_2021, gusev_viscous_2018,ginzburg_superballistic_2021}, large negative magnetoresistance\footnote{The viscous origin of negative magnetoresistance in high purity samples is disputed~\cite{gornyi_two-dimensional_2023}, therefore magnetic field dependence of viscosity is an open question.}~\cite{gusev_viscous_2018-1,shi_colossal_2014,keser_geometric_2021,mani_size-dependent_2013, alekseev_negative_2016,wang_hydrodynamic_2022,levin_geometric_2023},
Poiseuille flow profiles \cite{sulpizio_visualizing_2019,ku_imaging_2020,ella_simultaneous_2019}, Hall viscosity \cite{gusev_viscous_2018,berdyugin_measuring_2019},
negative non-local resistance and formation of whirlpools \cite{govorov_hydrodynamic_2004,kaya_nonequilibrium_2013,bandurin_negative_2016,braem_scanning_2018}, the development of local viscous regions \cite{samaddar_evidence_2021}, the violation of the Wiedemann-Franz
law\footnote{Although recent work in Ref.~\cite{tu_wiedemann-franz_2023} suggests that the result of a well-known experimental work \cite{crossno_observation_2016} is not hydrodynamic in nature.} \cite{crossno_observation_2016,gooth_thermal_2018,lucas_electronic_2018, jaoui_thermal_2021, ahn_hydrodynamics_2022}, resonant photoresistance in magnetic fields~\cite{dai_observation_2010,hatke_giant_2011,bialek_photoresponse_2015,alekseev_transverse_2019,wang_hydrodynamic_2022}, anomalous scaling of the resistance with the channel width \cite{gooth_thermal_2018,moll_evidence_2016}, and a quantum-critical dynamic conductivity~\cite{gallagher_quantum-critical_2019}. Moreover, novel phenomena have been predicted, such as anisotropic fluids~\cite{varnavides_electron_2020,link_out--bounds_2018}, the elimination of the Landauer-Sharvin resistance~\cite{stern_how_2022}, dynamo effect in the electron-hole plasma~\cite{galitski_dynamo_2018} and the possibility of hydrodynamic spin transport~\cite{tatara_hydrodynamic_2021,doornenbal_spinvorticity_2019,matsuo_spin_2020} inspired by experiments on liquid mercury~\cite{takahashi_spin_2016,takahashi_giant_2020}. In addition to the Poiseuille-like flow expected from electron hydrodynamics, anomalous flow profiles have been predicted in the boundary layer in charge neutral graphene in a uniform perpendicular magnetic field, with this ``anti-Poiseuille" flow attributed to recombination effects \cite{alekseev_counterflows_2018,alekseev_nonmonotonic_2018,narozhny_anti-poiseuille_2021}.

Hydrodynamic flow depends qualitatively on boundary conditions, with no-slip boundaries giving rise to the well known Poiseuille flow profile. On the other hand, it is also possible to achieve perfect-slip boundaries by using gates to create a channel in a clean semiconductor. This unique property was used in Ref.~\cite{keser_geometric_2021} for precise measurements of the viscosity in clean GaAs. The measurements indicate deviations from the predictions of the conventional theory of electron-electron scattering and this exciting observation requires more careful studies.

There exists a generic problem in electron hydrodynamics, namely disentangling Ohmic effects, boundary effects, and hydrodynamic effects in experimental data.
Viscous dissipation is proportional to the vorticity of the fluid, while the Ohmic dissipation 
is independent of vorticity. To enhance hydrodynamics one thus needs to enhance the vorticity. This might be achieved by generating turbulent flow of the electron fluid, but the electron fluid is extraordinarily viscous and therefore it is practically impossible to approach turbulence. 

In the laminar regime in a uniform magnetic field, vorticity is generated 
only near the boundaries of the channel in a layer of width $\sim\sqrt{l_{ee}l_{\rm mfp}}$. 
The flow in this boundary layer is influenced not only by the viscous fluid effects,
but also by the boundary conditions that are device specific and typically not well known. Therefore it is hard
to disentangle the unknown boundary effects and unknown viscous effects.

In Ref.~\cite{keser_geometric_2021} the problem was partially resolved by using a gated channel in 
a GaAs 2DEG that supports perfect-slip boundary conditions. This allowed for relatively accurate measurements of the viscosity of the electron fluid. However, the accuracy of the measurements was limited by Ohmic dissipation which was more than an order of magnitude larger than the viscous one.

In Ref.~\cite{engdahl_micromagnets_2022} we suggested a method that involved the use of micromagnets to create wiggling hydrodynamic flow and hence
create vorticity throughout the bulk of the sample, not just in the boundary region. This allows for a dramatic enhancement of viscous hydrodynamic dissipation.
Hydrodynamics in a finite channel is inherently a boundary problem and in Ref.~\cite{engdahl_micromagnets_2022} the problem was
solved only for a straight channel in the unique case of perfect-slip boundaries. In the present work we extend the method to a straight channel with arbitrary boundary conditions, either no-slip or finite-slip boundaries. 
Of course the bulk solution is the same, but the structure of the viscous boundary layer depends on the boundary condition and
we determine the structure in the present work. Hence, the present work extends the magneto-hydrodynamic method to graphene and also to $\delta$-doped semiconductors. The results of this work allow for the viscosity of the electron fluid to be extracted from a measurement of longitudinal resistivity of a finite straight channel in such devices.

The structure of this paper is as follows. In Sec.~\ref{sec:bulk_sol} we describe the solution for the bulk flow through a sinusoidal magnetic field found in Ref.~\cite{engdahl_micromagnets_2022}. The difference from Ref.~\cite{engdahl_micromagnets_2022} is that
we present numerical estimates for graphene instead of GaAs. In Sec.~\ref{sec:boundary_layer} we
study the boundary layer in a straight channel using no-slip
boundary condition and characterize the flow near the boundary. We first solve the boundary conditions using perturbation theory and then expand this idea to develop a numerically exact solution. The key results of Sec.~\ref{sec:boundary_layer} are the width of the boundary layer and the emergence of pseudo anti-Poiseuille flow at high magnetic field strengths. Understanding the width of the boundary layer allows a lower limit to be placed on the sample size in order to use the bulk solution for the ``zigzag'' flow. In Sec.~\ref{sec:num} we present numerical solution of the model developed in the previous section and show plots of dissipation and velocity profile.  In Sec.~\ref{sec:finite} we complement our previous numerical solution with a conventional finite element model and qualtitatively reproduce the flow profile. In Sec.~\ref{sec:magnets} we calculate the magnetic field of periodic ferromagnetic stripes
and hence determine parameters of micromagnets necessary for the experimental
set up assuming monolayer graphene with experimentally realistic density and mobility. 
Sec.~\ref{sec:concl} presents our conclusions.

	\section{Bulk Solution for Electron Hydrodynamics in a sinusoidal magnetic field} \label{sec:bulk_sol}
	We consider a channel that is sufficiently wide such that boundary layers are much thinner than the channel width, allowing for the bulk flow to be analysed indendently of boundary conditions. In this regime the fluid motion of the bulk may be analysed in the limit of an infinitely wide channel.
	
	 Consider a two-dimensional channel where $x$ is the longitudinal direction, $y$ the transverse direction and $z$ is perpendicular to the plane. The electron fluid may be controlled through the Lorentz force and thus obeys the stationary Navier-Stokes equation at constant density~\cite{landau_fluid_2013},
	\begin{align}
		\label{NS}
		\frac{ \mathbf{v}}{\tau}
		-\nu \nabla^2 \mathbf{v} 
  +\mathbf{v}\cdot{\nabla} \mathbf{v}
  &= -\frac{\nabla \Phi}{m^*} + \frac{q}{m^*} \mathbf{v}\times \mathbf{B}, \nonumber\\
		\nabla \cdot \mathbf{v} &=0.
	\end{align}
	Here, $\mathbf{v}$ is the fluid velocity field, 
	$\tau$ is the relaxation time related to momentum non-conserving scattering from
	impurities and phonons, $\nu$ is the kinematic viscosity, $\Phi$ is the electro-chemical potential, $m^*$ is the effective hydrodynamic
	mass, $q$ is
	the fluid particle charge and $\mathbf{B}$ is the magnetic field. For a uniform field, it is known that this equation predicts the usual Hall effect, whereas recently it has been realized that non-uniform fields have remarkable consequences such as a local suppression of flow~\cite{keser_magnetic_2023} and vortex layers in the bulk~\cite{engdahl_micromagnets_2022}.
The electron-electron scattering length, $l_{ee}$, and the momentum diffusive scattering length, $l_{\rm mfp}$,
are defined by the following relations
\begin{eqnarray}
\label{mfp2}
&&\nu=\frac{l_{ee}v_F}{4}\nonumber\\
&&\tau=\frac{l_{\rm mfp}}{v_F},
\end{eqnarray}
where $v_F$ is the Fermi velocity at zero temperature.

  The Navier-Stokes Eq.(\ref{NS}) is valid in the limit when $l_{ee}$ is much smaller than all geometric
  scales in the problem. There are two kinds of subleading corrections to the Navier-Stokes Eq.
  \cite{alekseev_negative_2016,scaffidi_hydrodynamic_2017}. The correction of the first kind, the dependence of
  viscosity on
  magnetic field, is parametrically proportional to $(l_{ee}/r_c)^2$, where $r_c=mv_F/(eB)$ is the cyclotron
  radius and the correction of the second kind, the Hall viscosity,  is parametrically proportional to
  $(l_{ee}/a)^2$, where $a$
  is the characteristic geometric scale. We start from the correction of the the second kind which
  leads to the following modification of the Navier-Stokes equation
	\begin{eqnarray}
		\label{NSm}
		&&\frac{ \mathbf{v}}{\tau}
		-\nu \nabla^2 \mathbf{v} 
                +\mathbf{v}\cdot{\nabla} \mathbf{v}=\left(1+\frac{l_{ee}^2}{2}\nabla^2\right)\mathbf{F}\\
&&\mathbf{F}=-\frac{\nabla \Phi}{m^*} + \frac{q}{m^*} \mathbf{v}\times \mathbf{B}\nonumber
	\end{eqnarray}
        The term $  (l_{ee}^2/2)\nabla^2$ acting on the Lorentz force,
        $(q/m^*) \mathbf{v}\times \mathbf{B}$, is literally the ``Hall viscosity''
        term from Ref.\cite{alekseev_negative_2016}.
        The term $ (l_{ee}^2/2)\nabla^2$ acting on the gradient of electrochemical potential,
        $-\nabla \Phi/m^*$,
        can be easily derived using the method used in Ref.\cite{alekseev_negative_2016}.
        The electrochemical potential is a sum of electrostatic part and the pressure part,
        $\Phi=\Phi_{E}+\Phi_{pr}$. Strictly speaking the  $ (l_{ee}^2/2)\nabla^2$ correction
        acts only on the electrostatic part. Since $\Phi_E \gg \Phi_{pr}$ we disregard this very small
        difference.
        Note that the term $ (l_{ee}^2/2)\nabla^2 \mathbf{F}$ is time even unlike the dissipative
       viscosity term $\nu \nabla^2 \mathbf{v} $ which is time odd. As this modified equation is the  Navier-Stokes plus the leading correction
       towards the kinetic equation, we call Eq.(\ref{NSm}) the NS$^+$ equation.


		A superlattice of micromagnets modulated in the $x$ direction generates a sinusoidal magnetic field along the $z$ direction in the plane of the 2DEG, 
  \begin{eqnarray}
      \label{Bz}
      B_z = B_0 \sin(gx) \ ,
  \end{eqnarray}
 where $g = 2 \pi/a$ and $a$ is the separation between adjacent micromagnets. 
 As previously mentioned, the electron fluid is very viscous, so the ``Bernoulli term'', $\mathbf{v}\cdot{\nabla} \mathbf{v}$, in the left hand side of Eq.~(\ref{NSm}) can be neglected.  This is because compared to the diffusion term $-\nu\nabla^2\mathbf{v}$ the Bernoulli term scales with the Reynolds number Re $=vL/\nu$, where $L$ is a characteristic length of the system. In this case the characteristic length scale is defined by the gradient operator, so an appropriate estimate is $L\approx 3D$ and upon subsituting in appropriate parameters later defined in Eq.~\eqref{param} we see Re$\approx10^{-3}$.  Hence, we obtain a linear NS$^+$  equation that can be solved
  using the stream function $\psi$ defined by $v_x =\partial_y\psi$ and $v_y =-\partial_x\psi$.
  For an infinitely long and infinitely wide channel with fixed averaged current density $\mathbf{J} = q n v_0 \mathbf{\hat{x}}$  the solution is
\begin{eqnarray}
\label{pinf}
\psi_\infty &=& v_0 \left( y +\frac{C\omega \tau}{g (1 + D^2 g^2)} \cos (gx)\right),
\nonumber\\
C&=&1-g^2l_{ee}^2/2
\end{eqnarray}
	where $D=\sqrt{\nu\tau}=\frac{1}{2}\sqrt{l_{ee}l_{\rm mfp}}$ is the characteristic viscous length and $\omega=|e|B_0/m^*$ is the cyclotron frequency. The viscous and Ohmic dissipation rates follow directly from Eq.~\eqref{pinf}, were $L$ and $W$ are the length and the width of the channel, respectively,
 \begin{eqnarray}
  \label{diss}
 \dot{E}_\Omega &=&\frac{nm^*}{\tau}\int d^2r\: v^2
 ,\nonumber \\
  \dot{E}_\nu &=&-nm^*\nu\int d^2 r\:\mathbf{v}\cdot \nabla^2 \mathbf{v}
\end{eqnarray}
Note that generally for the incompressible Navier-Stokes equation the expression for viscous dissipation may be explicitly split up into a bulk stress term and a boundary stress term\cite{landau_fluid_2013}, however for a periodic driving force the boundary term vanishes over an integer number of periods and vanishes when the flow at the inlet and outlet becomes uniform. The boundary stress term is also negligible when the channel length $L\to\infty$. We consider the long channel limit here, but it is also simple to ensure the other two cases are met via design of the device.  For the bulk solution (\ref{pinf}) Eqs.\eqref{diss} give
\begin{eqnarray}
  \label{diss1}
 \dot{E}_\Omega &=&\frac{nm^*v_0^2}{\tau}\ \left(1+  \frac{C^2\omega^2 \tau^2}{2(1+D^2 g^2)^2}
  \right) L W,\nonumber \\
  \dot{E}_\nu &=&  \frac{nm^*v_0^2}{\tau}\  \frac{C^2g^2D^2 \omega^2 \tau^2}{2(1+D^2 g^2)^2} L W.
\end{eqnarray}
  These Eqs. differ from a simlar solution in Ref.~\cite{engdahl_micromagnets_2022}
      by account of the leading gradient expansion correction, $C=1 \to C= 1-g^2l_{ee}^2/2$.
      Eqs.(\ref{diss1})  determine longitudinal resitance of a long and sufficiently wide channel.
      If the channel has about  50 micromagnets along the length  then the dissipation in the contacts is
      negligible compared to the bulk dissipation (not more than  1-2\%) and such a channel is said to be
      sufficiently long. The sufficient width
      is a more delicate issue and this
      question is addressed in Sec.~\ref{sec:boundary_layer} where we discuss the solution for the boundary
    layer. 
    Hence we propose to  measure the longitudinal resistance of a sufficiently long and sufficiently wide
    conducting channel.
      The diffusive mean free time $\tau$ can be can be determined from the measurement at zero magnetic field,
      Further measurements of the dependence of the resistance on the amplitude of the
      magnetic field allows the extraction of $l_{ee}$ and hence allows one to determine the viscosity $\nu$. Explicitly, this is achieved by taking the ratio of the total resistivity and the Drude resistivity,
      
\begin{equation}
	\frac{\rho_{Tot}}{\rho_D} = \frac{ \dot{E}_\Omega +  \dot{E}_\nu}{ \dot{E}_\Omega(\omega=0)} = 1+\frac{C^2\omega^2\tau^2}{2(1+D^2g^2)}.
\end{equation}

So far we have not explicitly solved for the electro-chemical potential in the Stokes equation~\eqref{NS}. Substituting $\mathbf{v}$ calculated from the stream function~\eqref{pinf}, it is easy to derive an expression for the electro-chemical potential, representing the external force on each fluid particle,
\begin{eqnarray}
    &&\partial_x \Phi = -\frac{m^* v_0}{\tau}\left( 1 + \frac{C^2\omega^2\tau^2 \sin^2(g x)}{1+D^2g^2} \right), \nonumber\\
    &&\partial_y\Phi = 0.
\end{eqnarray}
The electro-chemical potential does not vary in $y$ in an infinite sample as there is translational symmetry in the $y$ direction, whereas the periodic structure of the magnetic field in the $x$ direction gives rise to a modulated electro-chemical potential as a function of $x$. Furthermore, it is clear that the rate of work done by the external force $\dot{E} = -\int d^2 r\: \mathbf{v} \cdot\nabla \Phi $ matches the total dissipation $ \dot{E}_\Omega + \dot{E}_\nu $ obtained in Eq.~\eqref{diss1}. 

As the electron number density is approximately constant throughout the sample\footnote{See discussion leading to and following Eq.~\eqref{eq:incomp}}, the thermodynamic pressure gradients vanish. Therefore, the electrochemical potential gradients are predominantly electric in nature, and we can write the voltage as $\varphi = -\Phi/|e|$,
\begin{eqnarray}
    \label{eq:long_hall_eff}
    \varphi =  \frac{m^*v_0}{|e|\tau}\left(x 
 + \frac{C^2\omega^2\tau^2}{2(1+D^2g^2)} \left(x - \frac{\sin(2gx)}{2g}\right) \right).
\end{eqnarray}
From Eq.~\eqref{eq:long_hall_eff} it is clear that the electric field varies with $x$. As this effect arises from the presence of a periodic magnetic field and the modulation in voltage is in the same direction as transport it is appropriate to label this a ``magnetic field driven  longitudinal voltage modulation''. 

 Electron transport in high-mobility graphene may be hydrodynamic in the temperature range 
 $150\:\mathrm{K} \leq T <300\:\mathrm{K}$~\cite{narozhny_hydrodynamic_2022}. 
 To be specific we will present numerical estimates for a monolayer graphene device with mobility $\mu \approx 2\times10^5$ cm$^2$$/$Vs and density
 $n \approx 10^{12}$ cm$^{-2}$ at temperature $T = 220$ K with parameters similar to the device used in Ref.~\cite{bandurin_negative_2016},
 \begin{eqnarray}
 \label{param} 
 &&n=10^{12}\: \text{cm}^{-2}\nonumber\\
 &&T=220\:\text{K}\nonumber\\
 &&l_{\rm mfp}=1.9\:\mu \text{m}\nonumber\\
 &&l_{ee}=0.24\:\mu \text{m}\nonumber\\
 &&\tau=1.9\:\text{ps}
  \end{eqnarray}
For a magnetic field amplitude $B_0=75$ mT this gives $\omega\tau=1.2$.
   Here we keep in mind that the hydrodynamic mass for monolayer graphene is $m^* = p_F/v_F$, with Fermi momentum $p_F = \hbar \sqrt{n \pi}$, accounting for spin and valley degeneracy in graphene.
   With these parameters the characteristic viscous length is $D \approx 0.34\:\mu\mathrm{m}$.

Ohmic dissipation (\ref{diss1}) is dependent on the
dimensionless parameters $\omega\tau$ and $gD$ which, for fixed $\omega\tau$ ,
may be shown to be maximized at $gD=[1+(C\omega\tau)^2/2]^{1/4}$.
We take  $g=2.6$ $\mu$m$^{-1}$ which is slightly smaller than the optimal value
and which corresponds to lattice spacing $a=2.5$ $\mu$m. 
At a larger value of g (smaller a) the gradient non-Stokes correction
$l_{ee}^2\nabla^2/2\to l_{ee}^2g^2/2$ is too large. At $a=2.5$ $\mu$m the
correction is $l_{ee}^2g^2/2=0.19$.

With these parameters and assuming $v_0 = 100$ m/s, corresponding to
$I=0.15$ $\mu$A per $\mu$m of channel width  from
Ref.~\cite{bandurin_negative_2016}, the oscillation amplitude of the longitudinal voltage modulation is approximately $0.3$ $\mu$V. 
Of course the amplitude can be increased by increasing the current.

So far in the analysis we have assumed the electron fluid is incompressible
as the number density is made approximately constant via a gate voltage.
To check the validity of this assumption let us compare the gate voltage $\varphi_g$
that determines the electron density with the longitudinal voltage drop
$\varphi$.
\begin{equation} \label{eq:incomp}
\varphi_g \approx \frac{ned}{\epsilon}, \ \ \ \  \varphi \sim \frac{m^*v_0}{|e|\tau} L
\end{equation}
Here $d$ is the depth from the gate to the 2DEG, $\epsilon$ is the dielectric constant and
$L$ is the length of the device between Ohmic contacts,
Taking parameters from \eqref{param} and assuming a single sided hBN substrate with dielectric
constant $\epsilon_{hBN}=4.4$\cite{bandurin_negative_2016} such that $\epsilon = 2.7 $
and assuming $d=50$ nm, the gate voltage may be estimated as $\varphi_g \approx 3.4$ V.
For channel length $L=100$ $\mu$m the voltage drop along the device is $\varphi \approx 0.6$ mV.
Hence $\delta n/n \sim \varphi/\varphi_g \sim 2\times 10^{-4}$. The incompressible approximation is valid.

	\section{Boundary layer for a straight semi-infinite channel with no-slip boundary condition}
	\label{sec:boundary_layer}
              This section has two goals, (i) to estimate the width of the boundary layer, (ii) to
                point out some qualitative effects related to the boundary layer.
                In this section we disregard the subleading gradient correction and use  the standard
                Navier-Stokes equation (\ref{NS}).
                	
	\subsection{Boundary layer flow in the perturbative limit \texorpdfstring{$\omega\tau \ll 1$}{omega tau less than one}}
	\label{sec:pert_boundary_layer}
Consider a semi-infinite straight channel with a no-slip boundary at $y=0$ and fluid flow in the region $y>0$. In conjunction with the Stokes equation~\eqref{NS}, the stream function must satisfy the no-penetration and no-slip boundary conditions,
	\begin{equation}
		\label{bc}
		v_y|_{y=0}=-\partial_x\psi|_{y=0} = 0, \; v_x|_{y=0}=\partial_y\psi|_{y=0} = 0.
	\end{equation}
	The no-penetration condition ensures that no fluid can cross the boundary whereas the no-slip condition ensures that the fluid has zero tangential velocity at the boundary surface. Furthermore, the stream function must decay to the bulk solution~\eqref{pinf} far from the boundary.
	We look for a solution as an expansion in powers of $\omega\tau \ll 1$, i.e.,
	$\psi = \sum_{n=0}^{\infty} (\omega\tau)^n \psi^{(n)}$. First we solve for the zero-order solution, corresponding to flow in the absence of a magnetic field. Solving the Stokes equation for $\omega\tau = 0$ using the boundary conditions leads to the Poiseuille solution
	\begin{equation} \label{0th}
		\psi^{(0)} = v_0 \left(y + De^{-y/D}\right) .  
	\end{equation}
	Next, introducing the magnetic field as a perturbation we derive the first-order correction in the solution, $\psi = \psi^{(0)} + \omega\tau \psi^{(1)} $, 
	\begin{eqnarray} 
 \label{1st}
		\psi^{(1)} =  v_0 \cos (gx)\left( \frac{1}{g(1+g^2D^2)} + A e^{-y/D } \right. \nonumber \\+ \left. C_1 e^{-gy} + C_2 e^{-\sqrt{g^2+\frac{1}{D^2}}y} \right),
	\end{eqnarray}
where the coefficients $A$, $C_1$ and $C_2$ are defined as
\begin{align}
	 A &= \frac{1}{g(1-g^2D^2)}, \nonumber \\
	 C_1 &= \frac{2(g^2D^2+1)^{-\frac{1}{2}}-1}{g(1-g^2D^2)(gD-\sqrt{g^2D^2+1})}, \nonumber \\
	 C_2 &= \frac{1-gD}{g(1+gD)(1+g^2D^2)(gD-\sqrt{g^2D^2+1})}. \nonumber
\end{align}
Note that in the case of $g D\to 1$ the coefficients $A,C_1\to \infty$. However, it is easy to check that the divergences compensate each other such that the solution remains finite. 

To maximize the viscous dissipation we are interested in the case $gD > 1$. Since the slowest decaying boundary term in Eq.~\eqref{1st} decays as $e^{-y/D}$, the characteristic viscous length $D$ sets the width of the boundary layer.  
	
	\subsection{Exact numerical solution for the boundary layer flow}\label{sec:exact}
While the perturbative solution for the boundary layer presented in the previous subsection is illuminating, for practical purposes we must find the solution at $\omega \tau > 1$. This can only be done numerically.
To do so we reduce the Stokes equation~\eqref{NS} to an infinite series of linear algebraic equations. We then numerically solve these equations by truncating them at a suitable order.  

	\begin{figure*}[th!]
		\includegraphics[width=0.32\textwidth]{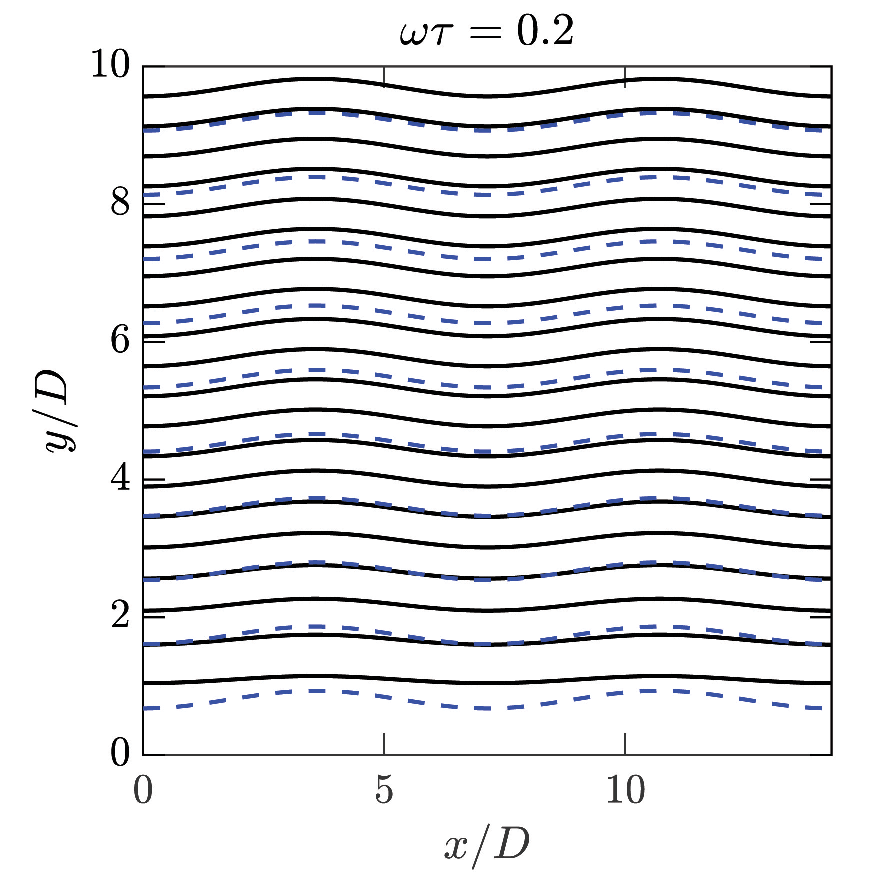}
		\includegraphics[width=0.32\textwidth]{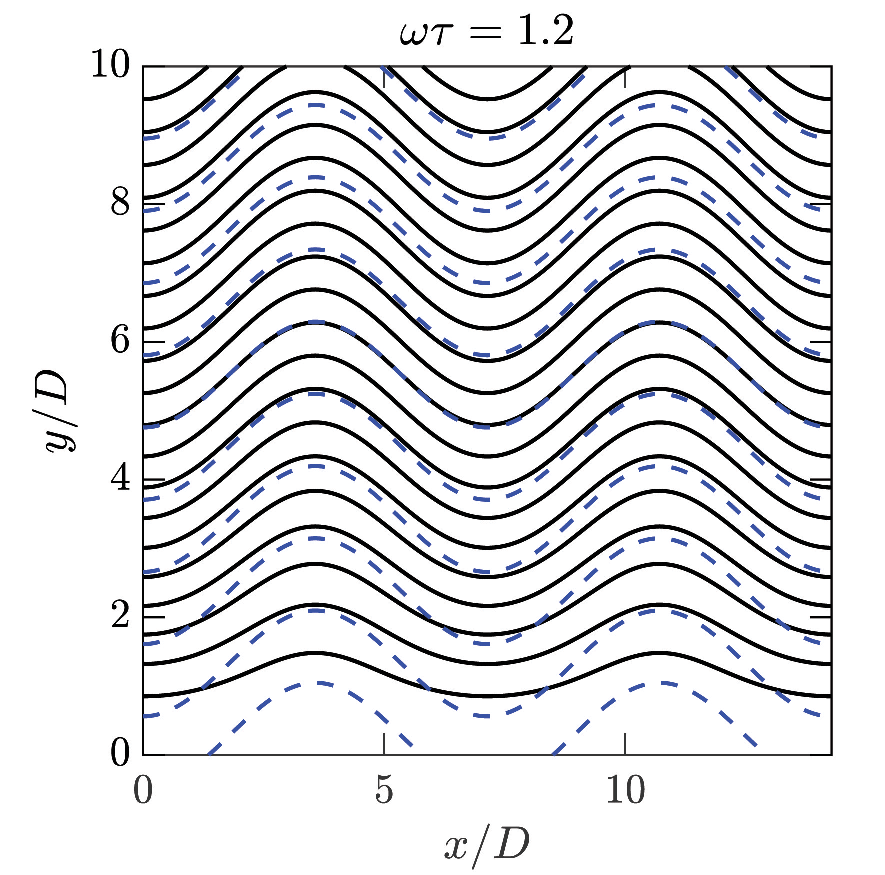}
		\includegraphics[width=0.32\textwidth]{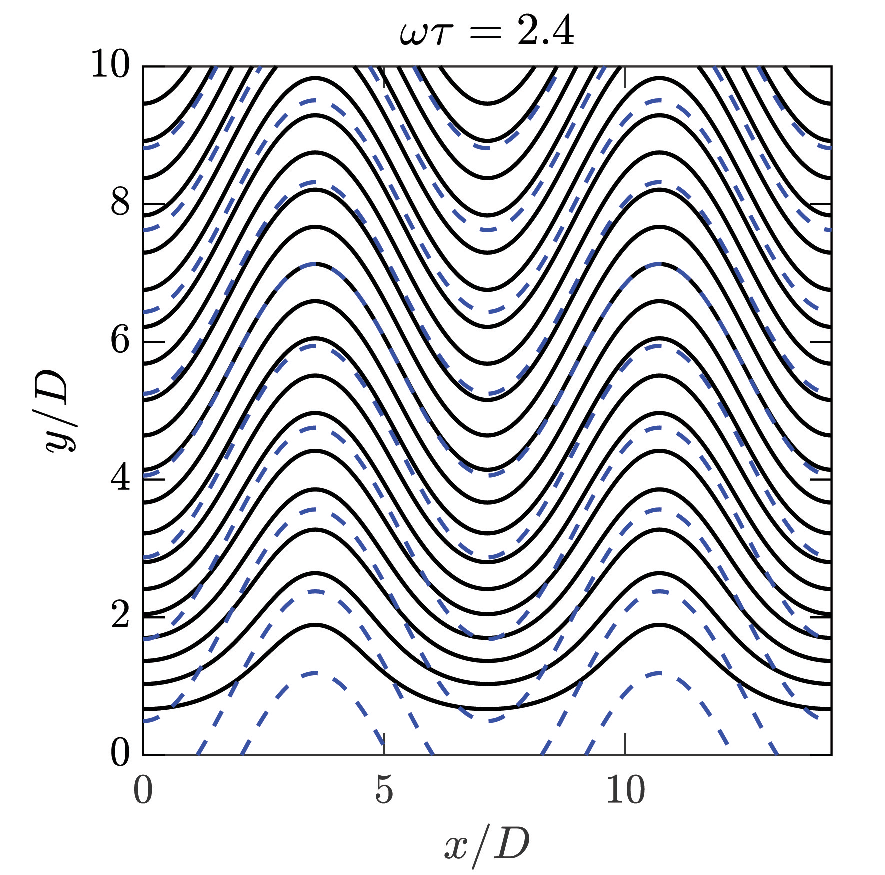}
		\caption{Streamlines in a semi-infinite channel with a no-slip boundary at $y = 0$. The parameters  are chosen as $gD=0.88$,
			$\omega\tau=0.2, 1.2, 2.4$.
			Solid lines correspond to the numerically exact solution described in
			Sec.~\ref{sec:exact} and dashed lines correspond to the solution given by Eq.~\eqref{pinf} in the
			infinite device.}
		\label{fig:exact}
	\end{figure*}

 We consider a semi-infinite channel $y\geq0$ and use length units where $D =\sqrt{\nu \tau}= 1$ and velocity units where $v_0 = 1$.
	As the magnetic field is periodic in the $x$ direction, so is the stream function $\psi$, which can thus be presented as the following Fourier series,
	\begin{eqnarray} \label{num_gen}
		\psi =y + e^{-y} + \frac{\omega \tau}{g(g^2+1)}\cos(gx)  \nonumber\\ + \sum_{m=-\infty}^{\infty}e^{i g m x } G_m(y)\ .	
	\end{eqnarray}
 Here $m$ is an integer and
	$G_m(y)$ are unknown functions that must decay with increasing $y$ as the solution tends to the bulk solution far from the boundary. Thus, using the perturbative solution as a starting point we consider solutions of the form 
 \begin{equation}
 \label{Gm}
     G_m(y) = f_m e^{-y} + \sum_a A_{m,a} e^{-\lambda_{a}y},
 \end{equation}
  where $f_m$ and $A_{m,a}$ are complex coefficients. The index $a=1,2,3...$ enumerates ``eigenvalues''
  $\lambda_a$, which can be real or complex. The real parts of $\lambda_a$ are positive by design of Eq.~\eqref{Gm}.
    In practice the upper limit of the index $a$, $a_{max} = 2N_m+1$, is fixed by truncation.
Since the ``Bernoulli term'' is neglected, the Stokes equation~(\ref{NS}) is a linear equation. Let us consider the Poiseuille solution without magnetic field,
\begin{eqnarray}
\label{Poi}
\psi_0 = y + e^{-y} \ ,
\end{eqnarray}
as a base solution. Now, in a magnetic field the stream function has a driven part
correction, $\delta \psi_d$, driven by $\psi_0$, and a free part correction, $\delta \psi_f$, 
	\begin{align}
 \label{psiv}
	\psi &=\psi_0+\delta \psi_d+\delta \psi_f,\nonumber\\
	\delta \psi_d &= \frac{\omega \tau}{g(g^2+1)}\cos(gx) + \sum_{m=-\infty}^{\infty}e^{i g m x }f_m e^{-y} ,\nonumber\\
	\delta \psi_f &= \sum_{m=-\infty}^{\infty}e^{i g m x} \sum_a A_{m,a} e^{-\lambda_{a}y}
        \end{align}
The first term in $\delta \psi_d$ is driven by the first term in Eq.~(\ref{Poi}) and the second term in $\delta \psi_d$
is driven by the second term in Eq.~(\ref{Poi}).
Hence, the function $G_m(y)$ in Eq.~(\ref{Gm}) is a combination of a driven part and a free part.
For a linear differential equation the driven part is fully determined by the equation, but the coefficients in the free part are determined by the boundary conditions.

It is convenient to define the following differential operators
		\begin{align}
		{\hat D} &\equiv -\nabla^2 + \nabla^4,\nonumber\\
		{\hat d} &\equiv \omega \tau g \cos(gx) \partial_y.
	\end{align}
 Using these notations the Stokes equation can be rewritten as
	\begin{align}
 \label{NSdf}
	(\hat{D} -{\hat d}) \delta \psi_d &= \hat{d}\psi_0, \nonumber\\
	(\hat{D} -\hat{d}) \delta \psi_f &= 0
\end{align}
	Using the representation (\ref{psiv}) the driven equation is transformed to an infinite set of linear algebraic equations for the coefficients $f_m$	
		\begin{multline} \label{driven}
			(g^2m^2)(g^2m^2-1)f_m + \frac{\omega \tau g}{2}(f_{m+1} + f_{m-1})\\ = -\frac{\omega \tau g}{2} (\delta_{m,1} + \delta_{m,-1}).
		\end{multline}
The coefficients in this equation are real and even in $m$.
Therefore, the coefficients $f_m$ are also real and even in $m$.
For this reason only equations for non-negative $m$ are independent.
A numerical solution of Eq.~(\ref{driven}) is straightforward, with truncation beyond
$m > N_m$, resulting in $N_m+1$ independent equations.
  The second equation in Eq.(\ref{NSdf}), i.e., the free equation, is transformed to 
	an infinite set of linear algebraic equations for the coefficients $A_m$,
		\begin{align} 
  \label{eigenproblem}
		(g^2m^2-\lambda_a^2)(1 +g^2m^2-\lambda_a^2) A_{m,a} \nonumber\\
  + \frac{\omega \tau g}{2}\lambda_a( A_{m+1,a} + A_{m-1,a}) =0,
	    \end{align}
noting that $\lambda_a$ is independent of $m$. Again, only equations with non-negative m are independent and we truncate $m$ such that $m \leq N_m$. 
Hence the number of independent equations is again $N_m+1$.
The right hand side of Eq.~(\ref{eigenproblem}) is zero,
so for a nontrivial solution the determinant of the matrix must be equal to zero.
Hence, this is an eigenvalue problem but it is not a standard linear eigenvalue problem.
To solve the eigenvalue problem we calculate the determinant as a polynomial in $\lambda$
and numerically determine the roots of the polynomial. The polynomial order is $4(N_m+1)$. 
After we discard roots with non-positive real part we are left with $2N_m+1$ roots.
Interestingly, at a sufficiently large $\omega\tau$ there are complex roots. The coefficients in the polynomial are real, so complex roots always come in complex conjugate pairs. 
We enumerate $2N_m+1$ roots by the index $a$.
For any found $\lambda_a$ we solve Eqs.~(\ref{eigenproblem}) and hence find the
coefficients $A_{m,a}$, numerically determined to arbitrary normalization. We now write $A_{m,a}=\alpha_aA_{m,a}^{(n)}$, where $A_{m,a}^{(n)}$ is
 normalized such that $\sum_m |A_{m,a}^{(n)}|^2=1$, and $\alpha_a$ is an arbitrary coefficient.

The boundary conditions~\eqref{bc} provide a set of equations to determine the $2N_m+1$ coefficients $\alpha_{a}$		
\begin{eqnarray} 
  \label{nopen}
&&0=v_y|_{y=0} \Longrightarrow \nonumber\\
&&m(f_m + \sum_a \alpha_aA_{m,a}^{(n)}) + \frac{\omega \tau}{2g(1+g^2)}(\delta_{m,1}-\delta_{m,-1}) = 0, \nonumber\\
  && 0=  v_x|_{y=0}\Longrightarrow\nonumber\\
&&f_m + \sum_a \lambda_a \alpha_aA_{m,a}^{(n)} =0.
	\end{eqnarray}
The first of these equations is nontrivial at $m=1,2...N_m$ and the second equation is nontrivial
at $m=0,1...N_m$, so altogether we have $2N_m+1$ equations to determine
$2N_m+1$ unknown coefficients $\alpha_a$.
Thus the boundary layer problem is solved.

The boundary layer solution is extended to finite width channels in App.~\ref{app_2}.
After the method is developed it is instructive to present plots for specific cases.
For the numerical solution we set $gD=0.88$, which corresponds to the device described at the end of Sec.~\ref{sec:bulk_sol}.
We truncate beyond $N_m=5$, which provides more than sufficient accuracy for the values
of $\omega\tau$ we present.  Plots of the streamlines obtained for $\omega\tau=0.2$, $1.2$ and $2.4$ are presented in Fig.~\ref{fig:exact}. These values of $\omega\tau$ were selected as $\omega\tau=0.2$ has only real eigenvalues $\lambda_a$
and corresponds to the perturbative regime, $\omega\tau=1.2$ has a complex conjugate pair of eigenvalues $\lambda_a$ and corresponds to the device described in Eq.~\eqref{param} and $\omega\tau=2.4$ has two complex conjugate pairs of eigenvalues $\lambda_a$.
As the magnetic field increases ($\omega \tau$ increases) both the magnitude of the boundary effects and
the width of the boundary layer increase. For the most interesting case $\omega\tau \approx1 $ the effective width of the boundary layer
  is approximately $D$. Therefore, for the set of parameters  (\ref{param}) a channel of width $=10-20\mu$m
is suffiently wide to neglect boundary layer effects.  

The  boundary layer solution considered in this Section is valid for the no-slip boundary
condition. However, the method can be easily extended to the case of an arbitrary slip length, see App.~\ref{finite_slip} for details.
 
\section{Numerical estimates, anti-Poiseuille flow} \label{sec:num}
\begin{figure}[ht!]
	\includegraphics[width=0.5\textwidth]{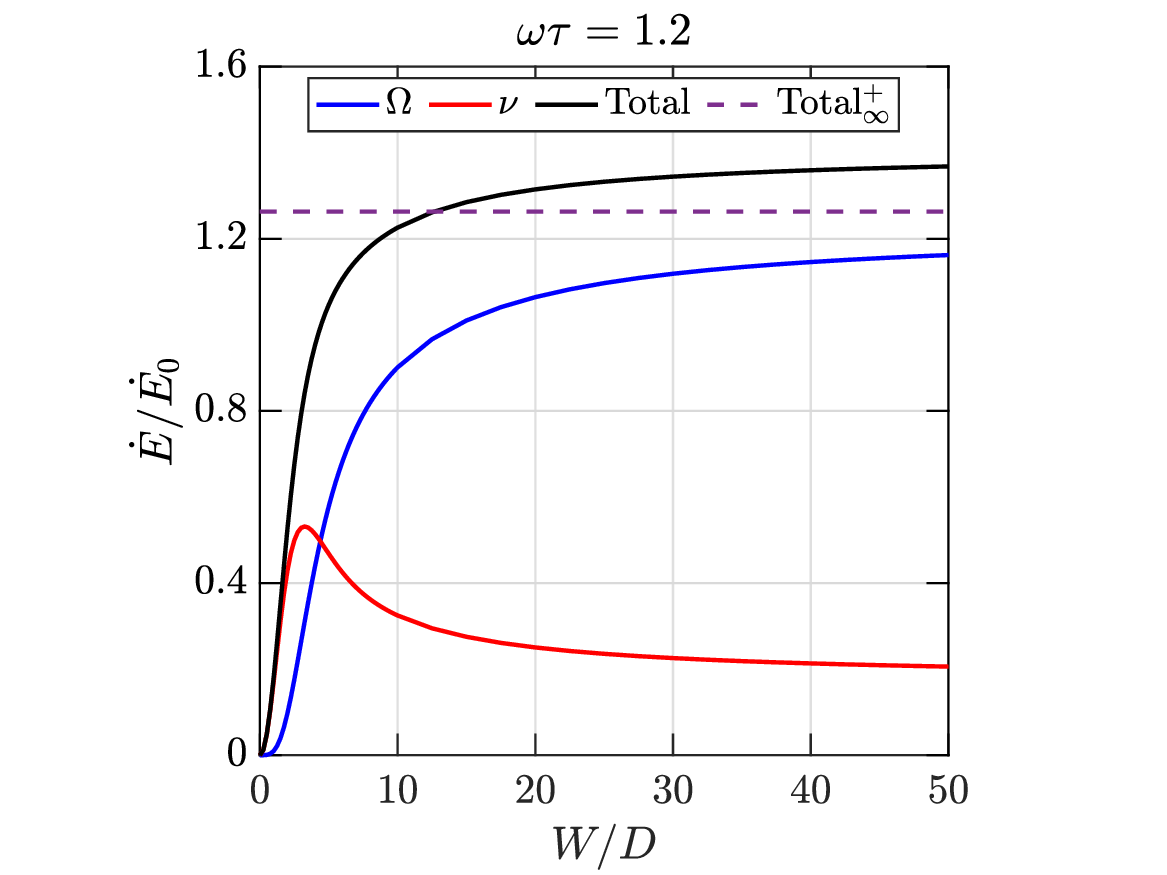}
	\includegraphics[width=0.5\textwidth]{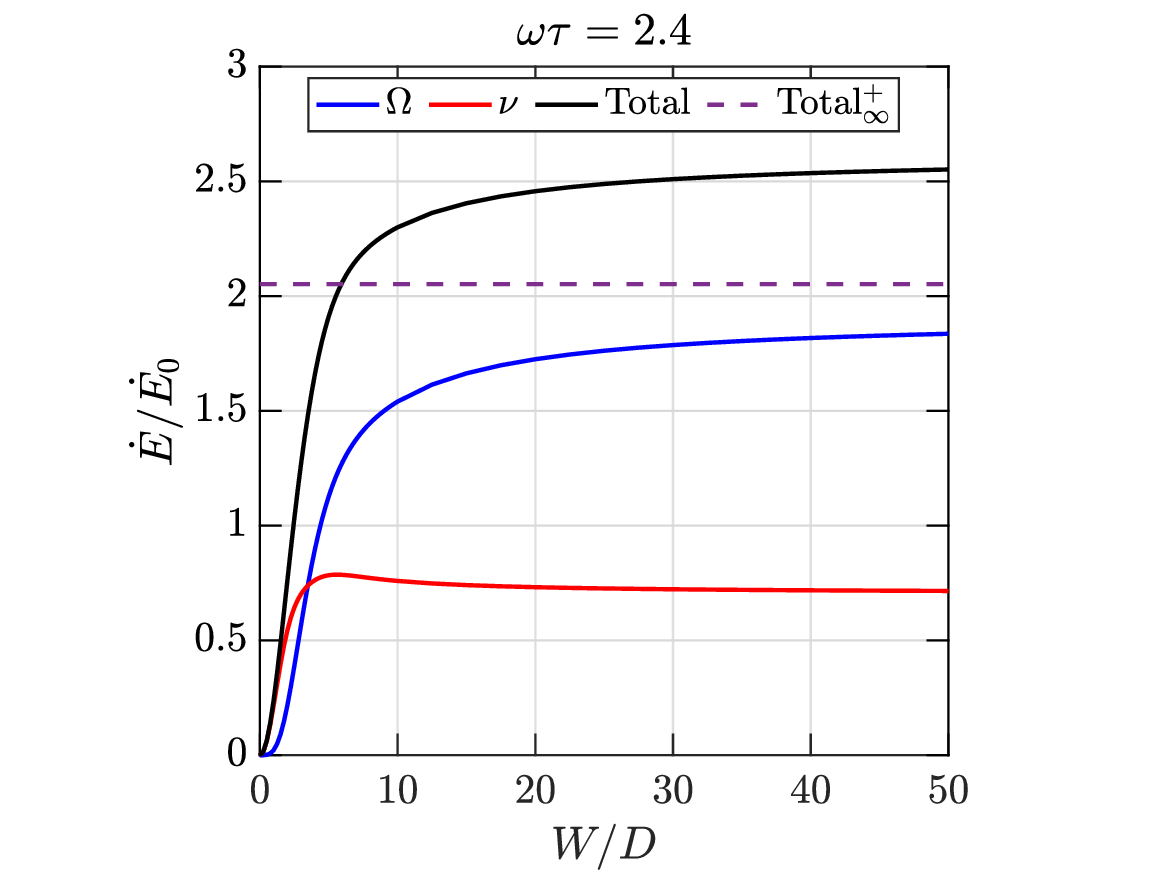}
	\caption{Ohmic, viscous and total dissipation from solution of the NS equation~\eqref{NS} as a function of channel width and total dissipation in the bulk from solution of the NS$^+$ equation~\eqref{NSm}. The dissipation is normalized to the $\dot{E}_0$ which is due to Drude resistance only, i.e., what is expected in the absence of B-field and hydrodynamics ($\dot{E}_0/LW =nm^*v_0^2/\tau$). The value $\omega\tau=1.2$ corresponds to $B_0=75$ mT and $\omega\tau=2.4$ corresponds to $B_0=150$ mT. In these plots we assume that the viscosity is independent of the magnetic field, $\nu(B)=\nu(0)$. The plots correspond to the parameter set (\ref{param}), where $g D = 0.88$.}
	\label{fig:e_dot}
\end{figure}

Following the numerical calculation of the stream function, we may calculate the
the viscous dissipation rate and the Ohmic dissipation rate given by Eqs.~(\ref{diss1}).
The most important parameter is the ratio $\dot{E}_\nu/\dot{E}_\Omega$, which is dependent on the device considered, so to provide calculated values we take the device with parameters presented in Eq.~(\ref{param}).
This corresponds to $D=0.34\:\mu\text{m}$. We also take the  lattice spacing $a=2.5\:\mu\text{m}$ as discussed in the end of Section \ref{sec:bulk_sol}, resulting in $gD=0.88$.
The dissipation rates are plotted in Fig.~\ref{fig:e_dot} versus the channel width $W$ for two values of the magnetic field amplitude, $B_0=75$ mT, $150$ mT, corresponding to $\omega\tau = 1.2$, $2.4$ respectively. Solid lines correspond to solutions of the NS equation~\eqref{NS}. The dashed line corresponds to total dissipation in the bulk from solution of the NS+ equation~\eqref{NSm} where the gradient correction is significant, $C^2=1 \to C^2=(1-0.19)^2=1-0.34$. In these plots we assume that the viscosity is independent of the magnetic field, $\nu(B)=\nu(0)$. 
Asymptotic values of the ratio $\dot{E}_\nu/\dot{E}_\Omega$ at $W\to \infty$ are $0.16$ ($B_0=75$ mT)
and $0.48$ ($B_0=150$ mT). Note that as magnetic field increases the difference between the asymptotic value of the total dissipation from the NS equation and the total bulk dissipation from the NS$^+$ equation increases. 

Now let us calculate the value of the correction of the first
kind as it is discussed in the paragraph before Eq.\eqref{NSm}.
The viscosity is often reported to depend on magnetic
field, with possible dependence~\cite{alekseev_negative_2016}
\begin{align}
	\label{bstar1}
	\nu \to \frac{\nu}{1+(B/B^*)^2} &\approx\nu[1-(B/B^*)^2]
	\;,\; \nonumber\\B^*& = \frac{p_F}{2|e|l_{ee}}
\end{align}
The term  $\nu (B/B^*)^2 \nabla^2 \mathbf{v}$ added to the Navier-Stokes
Eq. \eqref{NS} leads to the correction of the first kind.
we call it "the amplitude correction".
It is easy calculate the correction by perturbation theory.
In perturbation theory it does not influence the Ohmic dissipation, but
in calculating the viscous dissipation in Eq.(\ref{diss}), as $B=B(x)$ gives $\nu=\nu(x)$ one has to move $\nu$
in the integrand and perform integration over position.
This gives the following correction to the viscous dissipation.
\begin{eqnarray}
	\label{ac}
	\frac{\delta {\dot E}_{\nu}}{{\dot E}_{\nu}}=
	-3(\omega\tau)^2\left(\frac{l_{ee}}{l_{mfp}}\right)^2  
\end{eqnarray}
The correction is $-7\%$ for $\omega\tau=1.2$ and $-28\%$ for $\omega\tau=2.4$.
Note that by performing experiments for different values of $a$ and
different values of the magnetic field amplitude it is possible to
disentangle and to measure independently the viscosity,
the gradient correction (Hall viscosity), and the amplitude correction (\ref{ac}).

\begin{figure}[ht!]
	\includegraphics[width=0.45\textwidth]{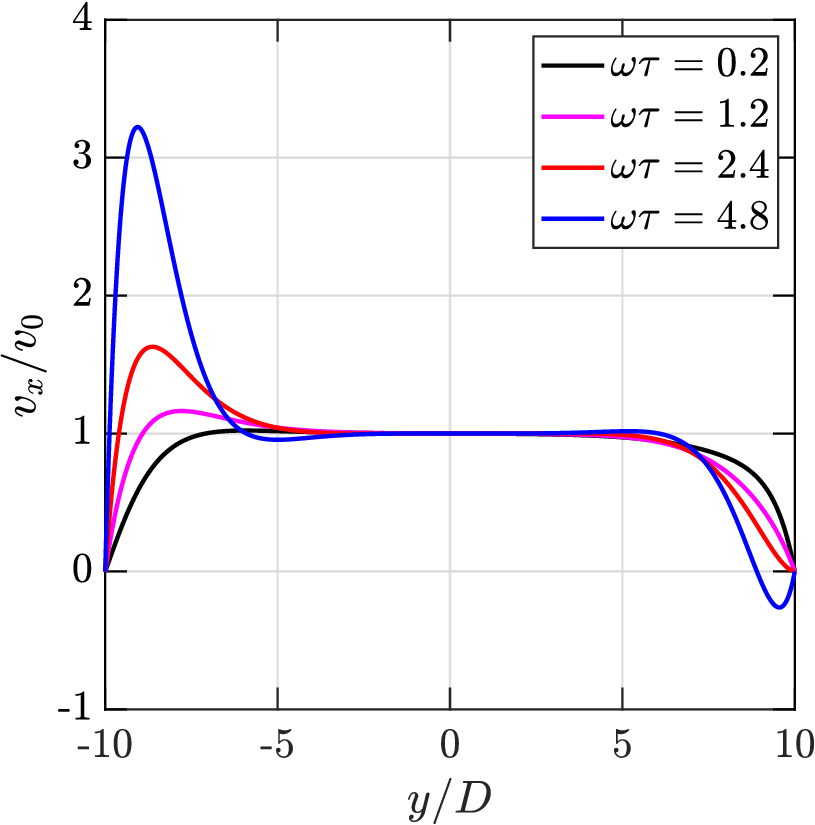}
	\caption{Longitudinal velocity profile at $x/D = 0$ with $gD=0.88$ for various values of $\omega\tau$. At large magnetic fields anti-Poiseuille flow is observed and a small counterflow is observed at the opposite boundary due to the formation of a viscous vortex.}
	\label{fig:v_x}
\end{figure}

Interestingly, the fluid flow in the boundary layer shows a sort of anti-Poiseuille behaviour.
To illustrate this, we plot in Fig.~\ref{fig:v_x} the longitudinal velocity $v_x$ versus $y$ (i.e., across the channel). The velocity $v_x$ is a function of both $x$ and $y$ and here we present plots at $x=0$. At a sufficiently large magnetic field, $\omega\tau \geq 1.2$, there is a clear anti-Poiseuille profile near the left boundary of the channel. In this regime the peak velocity is not $v_0$ in the bulk, but instead the longitudinal velocity has a maximum in the boundary layer before decaying towards $v_0$ in the bulk. 
There is a conventional Poiseuille behaviour at the right boundary of the channel, although at high magnetic field a small amount of counterflow develops due to the development of vortices in the boundary layer.  
If we shift $x=0 \to x=a/2$ the plots are effectively reflected about the centre of the channel: the anti-Poiseuille profile 
is located at the right boundary and the Poiseuille profile is at the left boundary.
Mathematically, the anti-Poiseuille profile is related to the complex eigenvalues $\lambda_a$.
 This behaviour is qualitatively similar to the anti-Poiseuille flow predicted for narrow graphene channels near charge neutrality in the presence of a uniform magnetic field perpendicular to the plane \cite{alekseev_counterflows_2018,alekseev_nonmonotonic_2018,narozhny_anti-poiseuille_2021}. Near charge neutrality this effect is largely attributed to electron-hole recombination.
 In our case the physical origin of the anti-Poiseuille profile is different as we do not have any holes.
  While we consider no-slip boundaries, a similar anti-Poiseuille profile exists also for boundary conditions with a finite slip
  length, see Appendix \ref{finite_slip}. 

\section{Finite-elements simulations} \label{sec:finite}

To complement our numerical results and to extend them towards finite systems, we also performed numerical simulations using the finite-element method (FEM). The first step is to bring Eq.~(\ref{NS}) to the form of a weak differential equation by multiplying each of the equation with a test function and integrating over all space. We again neglect the Bernoulli term because we assume small flow velocity and assume the fluid to be incompressible. For the finite-element analysis, we use the space of Lagrange polynomials of second order as the function space for the velocity field and the space of Lagrange polynomials of first order as the function space for the potential. Together these form a Taylor-Hood element appropriate for the numerical solution of Navier-Stokes equations. We have numerically implemented the problem using the \textit{FEniCS} package \cite{LangtangenLogg2017}.

Using FEM, we consider a finite system with a rectangular geometry with width $W$ in $y$ direction and length $L$ in $x$ direction. To simulate a finite array of micromagnets, we consider a flow driven by an applied electric potential difference, so we assume Dirichlet boundary conditions for the voltage $\varphi(x=-L/2,y) = \varphi_L$ and $\varphi(x=L/2,y) = \varphi_R$ at inlet and outlet. In addition to this, we assume that $v_y(x=\pm L/2,y) = 0$ at both inlet and outlet. Finally, we impose no-slip boundary conditions at the walls, i.e., $\mathbf{v}(x,y=\pm W/2) = 0$.

\begin{figure}[ht!]
	\includegraphics[width=0.5\textwidth]{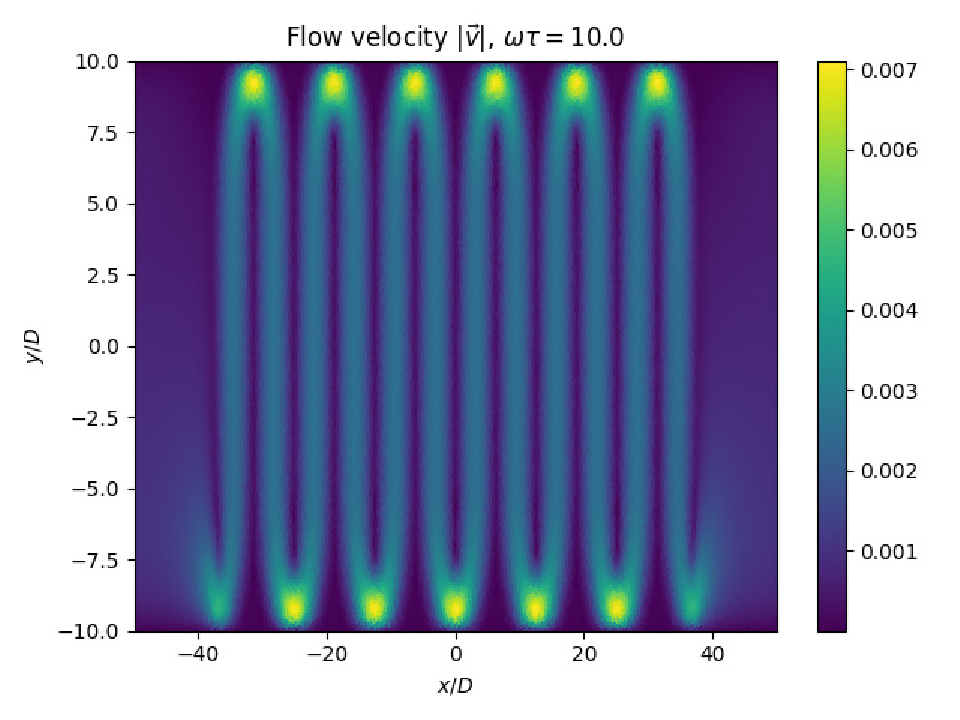}
	\caption{FEM results for the absolute value $|\mathbf{v}|$ of the velocity in a finite channel of length $L=100D$ for $gD = 0.5$. There is an equilibration region near the inlet and outlet where uniform flow transitions to periodic flow, with periodic flow established from $-40<x/D<40$. Note that this color plot should not be confused with lines of flow, which are shown in Fig.~\ref{fig:exact} The velocity is measured in units convenient for numerical solution and the system contains six magnetic strips. The region of high velocity, anti-Poiseuille flow, is shifted by half a period at opposite boundaries.}
	\label{fig:FEM_v}
\end{figure}

Figure~\ref{fig:FEM_v} shows that the results of these simulations qualitatively reproduce all features found for the infinite system. At strong enough magnetic fields, an anti-Poiseuille flow emerges near the boundary. Moreover, at strong fields, a small counterflow becomes visible at the opposite boundary.

\section{Magnetic field of periodic ferromagnetic stripes}
\label{sec:magnets}
\begin{figure}[t]
	\includegraphics[width=0.5\textwidth]{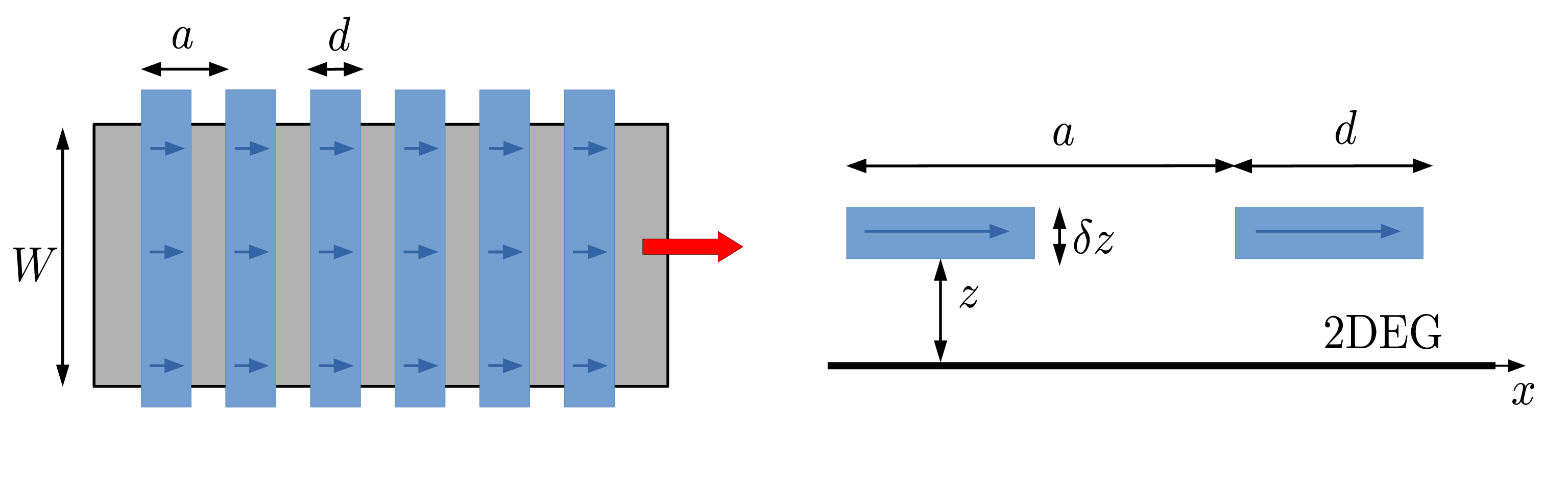}
	\caption{Left: top view of a periodic array of magnetic bars perpendicular to
		the flow direction shown by the red arrow. The micromagnets are magnetized parallel to the longitudinal axis of the channel. The direction of the magnetization is shown by the blue arrows. The period of the superlattice is $a$ and
		the width of each micromagnet is $d$.
		Right: Side view of the device. The vertical distance from the micromagnets to the 2DEG is $z$ and the
		thickness of each micromagnet is $\delta z$. 
		First presented in Ref.~\cite{engdahl_micromagnets_2022}.
	}
	\label{MB}
\end{figure}

The setup considered in this paper is similar to the setup previously proposed for GaAs heterostructure in Ref.~\cite{engdahl_micromagnets_2022}, but adapted to suit monolayer graphene with parameters described in Eq.\eqref{param}. A superlattice of rectangular magnetic bars oriented perpendicular to the longitudinal axis of the device is placed above the 2DEG as seen in the left panel of Fig.~\ref{MB}.

The thickness of each micromagnet  is $\delta z$ and the distance from the base of each micromagnet to the 2DEG is $z$,
shown in the right panel of Fig.~\ref{MB}.
The period of modulation is $a=2.5$ $\mu$m as found in
Sec.~\ref{sec:bulk_sol}. The micromagnets are magnetized parallel to the longitudinal axis of the channel. The remaining free parameters $d,z,\delta z$ are tuned such that
the $z$ component of the magnetic field in the plane of the 2DEG is approximately
$B_0\sin(2\pi x/a)$ with $B_0\approx 75$ mT. 
A stronger magnetic field may be also achieved.

Assuming the magnetization is uniform throughout the volume of each micromagnet, and each micromagnet's extent in the transverse ($y$) direction is wider than the width $W$ of the sample, we calculate the $z$ component of the magnetic field in the plane of the 2DEG as a function of longitudinal coordinate $x$ from the Biot-Savart law as
\begin{align}
	\label{Bfield}
	\frac{B_z(x)}{B_m}\bigg\vert_{\rm 2DEG} &= -\sum_{n>1} \frac{\sin\left(g n d/2\right)}{ \pi n} \sin(g n x) Z_n,\notag \\
	Z_n(z,\delta z)&= e^{-g n z} (1-e^{-g n \delta z}),
\end{align}
where $B_{m}$ is the saturation magnetic field, and $x$ is measured from the center of one of the micromagnets. At the edge of the micromagnet centred on $x=d/2$, the harmonics of the field in the 2DEG plane is $\propto \sin^2(g  n d/2)$, so evidently $d=a/2 = \pi/g$ maximizes the first harmonic and suppresses the second harmonic. We calculate $B$ numerically from Eq.~\eqref{Bfield} and confirm that $d=a/2=1.25\:\mu$m is optimal.

\begin{figure}[ht!]
		\includegraphics[width=0.48\textwidth]{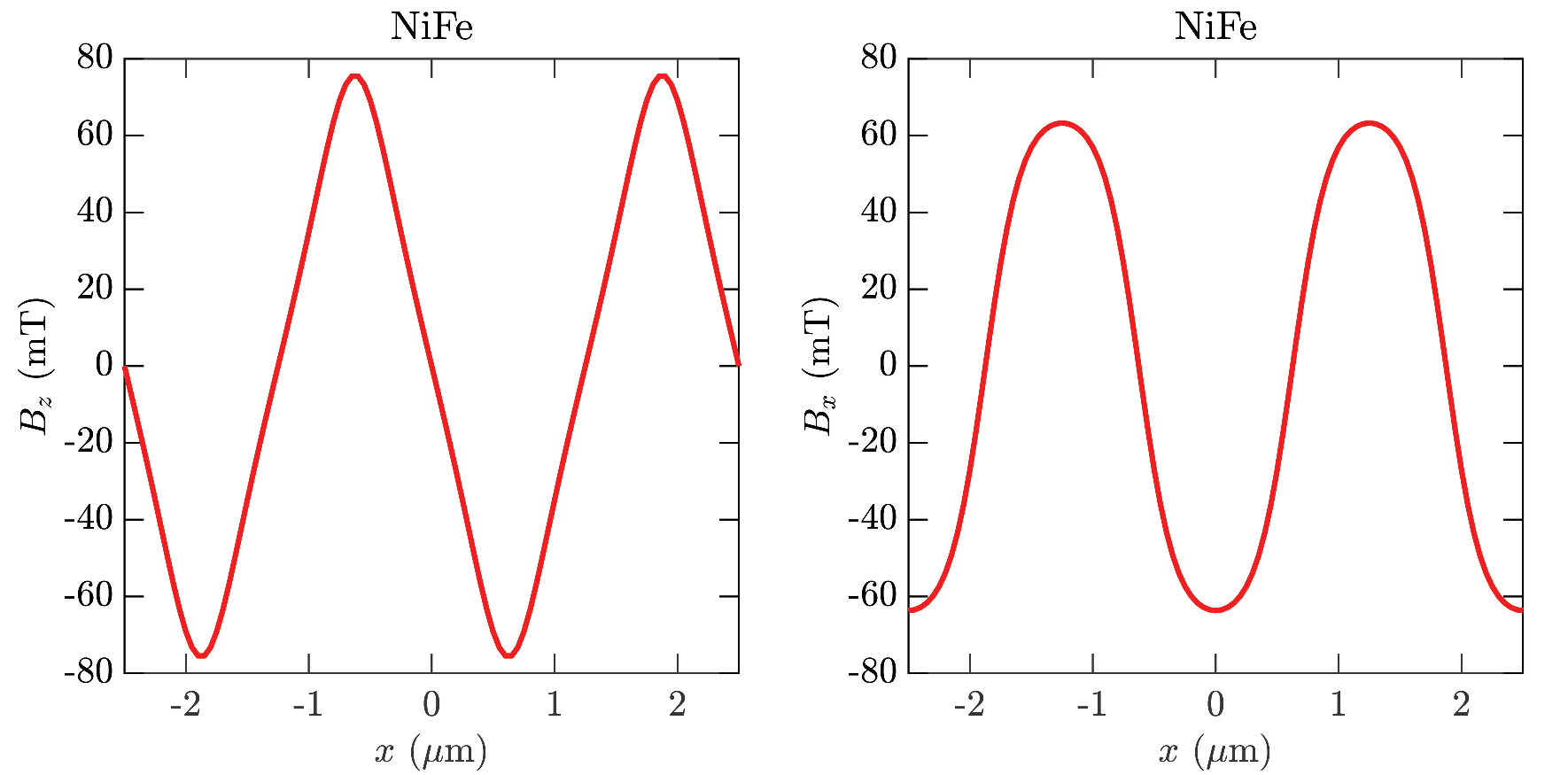}
\caption{ Magnetic field in the 2DEG plane for NiFe bars. Left panel:
			$B_z$ versus $x$. Right panel: $B_x$ versus $x$.
		The parameters are
		 $\delta z=0.20\:\mu\mathrm{m},  \ z=0.40\:\mu\mathrm{m}$.
	}
	\label{NiFe}
\end{figure}

In order to achieve a field with magnitude $B_0 = 75$ mT we consider a NiFe alloy that 
has saturation magnetic field $1500$ mT. \cite{tumanski_magnetic_2011}
Through concurrently tuning $\delta z$ and $z$ we find that we may generate the desired magnetic field profile with $\delta z = 0.20$ $\mu$m and $z = 0.40$ $\mu$m. Plots of $B_z$ versus $x$ and $B_x$ versus $x$ are presented in  Fig.~\ref{NiFe}.
The $x$ component of the magnetic field is not relevant, but we present it for completeness.
The $x$ dependence of $B_z$ is not quite sinusoidal, however this is not a big issue as
using the finite element method one can also solve the Navier-Stokes equation for this magnetic field profile. It should be noted that any magnetic disorder due to a slight misalignment of the micromagnets with respect to the channel or inconsistencies in the spacing or width of the micromagnets results in high wavenumber contributions to the magnetic field, with these contributions exponentially suppressed with due to the distance $z$ between the micromagnets and the 2DEG as seen in Eq.~\eqref{Bfield}.

\section{Conclusions} \label{sec:concl}
We develop a theory for the no-slip boundary layer in the magnetohydrodynamic flow of an electron fluid in a straight channel. The fluid flow is modified by an array of micromagnets on the top of two-dimensional electron gas. The micromagnets create a wiggling flow of the fluid, which
dramatically enhances hydrodynamic effects in the fluid. To be specific, the analysis was performed for no-slip boundary conditions,
but the method can be extended to boundary conditions with arbitrary slip length.
We demonstrate that the fluid velocity profile in the boundary layer manifests an anti-Poiseuille behaviour. In addition, we show that the micromagnet array creates a longitudinal voltage modulation in the sample. From the experimental point of view we propose a method for a boundary-independent measurement of the viscosity of different
electron fluids. The results are applicable to graphene away from the charge neutrality point and also to semiconductors.

\section*{ Acknowledgements} 
We acknowledge important discussions with Alexander Hamilton, Yonatan Ashlea Alava, Oleh Klochan, Daisy Wang and Zeb Krix. This work was supported by the Australian
Research Council Centre of Excellence in Future Low-
Energy Electronics Technologies (CE170100039). 

\appendix

\section{The method for exact solution for a finite width channel} \label{app_2}
Here we expand the method of Sec.\ref{sec:boundary_layer} to a finite channel with no slip boundaries at $y=\pm W/2$. Here we use units $D=v_0=1$. As before we split the stream function into the zero field term which corresponds to Poiseuille flow, $\psi_0$, the driven magnetic field correction term, $\delta \psi_f$ and the free magnetic field correction term $\delta \psi_d$. 
	\begin{align} \label{eq:app_psi}
	\psi &=\psi_0+\delta \psi_d+\delta \psi_f,\nonumber\\
        \psi_0 &= y - \frac{\sinh{y}}{\cosh{W/2}}\nonumber\\
	\delta \psi_d &= \frac{\omega \tau}{g(g^2+1)}\cos(gx) + \sum_{m=-\infty}^{\infty}e^{i g m x }\left(f_m e^{-y} + h_m e^{y}\right) ,\nonumber\\
	\delta \psi_f &= \sum_{m=-\infty}^{\infty}e^{i g m x} \sum_a \left( A_{m,a} e^{-\lambda_{a}y} + C_{m,a} e^{\lambda_{a}y}\right)
        \end{align}

Note that we absorb the constant exponential terms into the unknown coefficients. The stream function solves the Stokes' equation as in Eq.\eqref{NSdf}. From the driven equation we obtain a set of equations for coefficients $f_m$ and $h_m$. 

		\begin{align} \label{app_driven}
			(g^2m^2)(g^2m^2-1)f_m + \frac{\omega \tau g}{2}(f_{m+1} + f_{m-1})\\\nonumber = -\frac{\omega \tau g}{4\cosh{W/2}} (\delta_{m,1} + \delta_{m,-1})\\\nonumber
   			(g^2m^2)(g^2m^2-1)h_m - \frac{\omega \tau g}{2}(h_{m+1} + h_{m-1})\\\nonumber = -\frac{\omega \tau g}{4\cosh{W/2}} (\delta_{m,1} + \delta_{m,-1})
		\end{align}
As the coefficients are real and even in $m$, $f_m$ and $h_m$ are also real and even in $m$. The free equation gives eigenequations for eigenvalues $\lambda_a$ and eigenvectors $A_{m,a}$ and $C_{m,a}$.
		\begin{eqnarray} 
  \label{app_eigenproblem}
		(g^2m^2-\lambda_a^2)(1 +g^2m^2-\lambda_a^2)) A_{m,a} \nonumber\\
  + \frac{\omega \tau g}{2}\lambda_a( A_{m+1} + A_{m-1}) =0\nonumber\\
  		(g^2m^2-\lambda_a^2)(1 +g^2m^2-\lambda_a^2)) C_{m,a} \nonumber\\
  - \frac{\omega \tau g}{2}\lambda_a( C_{m+1} + C_{m-1}) =0
	    \end{eqnarray}
Note that the determinant of the $A$-equation is identical to that of the $C$-equation.
Hence, there is only one set of eigenvalues $\lambda_a$ which we find following the method described in the main text.
The  eigenvectors $A_{m,a}$ and $C_{m,a}$ are different and we find them solving Eqs. (\ref{app_eigenproblem}).
The eigenvectors are determined up to some arbitrary normalization. Similarly to the approach in
Sec.\ref{sec:boundary_layer}, we define coefficients $\alpha_a$ and $\gamma_a$ and write $A_{m,a}=\alpha_aA_{m,a}^{(n)}$ and $C_{m,a}=\gamma_aC_{m,a}^{(n)}$, where $\sum_m \left|A_{m,a}^{(n)}\right|^2=1$ and $\sum_m \left|C_{m,a}^{(n)}\right|^2=1$, and $\alpha_a$ and $\gamma_a$ are arbitrary coefficients. These coefficients are determined by considering the boundary conditions at $y=\pm \frac{W}{2}$. 

\begin{align}
  \label{app_nopen}
&0=v_y|_{y=\pm W/2} \to\nonumber\\
&m\bigg[f_m e^{\mp W/2} + h_m e^{\pm W/2} + \sum_a ( \alpha_aA_{m,a}^{(n)}e^{\mp\lambda_aW/2} \nonumber\\&+ \gamma_aC_{m,a}^{(n)}e^{\pm\lambda_aW/2} ) \bigg]+ \frac{\omega \tau}{2g(1+g^2)}(\delta_{m,1}-\delta_{m,-1}) = 0\nonumber\\
  &0=  v_x|_{y=\pm W/2}\to\nonumber\\
& \sum_a \lambda_a( \alpha_aA_{m,a}^{(n)}e^{\mp\lambda_aW/2}-\gamma_aC_{m,a}^{(n)}e^{\pm\lambda_aW/2}) \nonumber\\&+ f_m e^{\mp W/2} - h_m e^{\pm W/2} =0
\end{align}

Upon solving Eq.~\eqref{app_nopen} and obtaining $\alpha_a$ and $\gamma_a$ the stream function is entirely characterized and velocity and dissipation profiles may be calculated. 

\section{Solution for Boundaries With Finite Slip Length} \label{finite_slip}
We now expand upon the solution for the finite channel to account for some finite slip length, $\beta$, defined in the finite-slip boundary condition.\cite{kiselev_boundary_2019} Note the change in sign of $\beta$ at the positive and negative $y$ boundaries.
\begin{equation}
	v_x|_{y=\pm W/2} = \mp\beta \partial_y v_x|_{y=\pm W/2}
\end{equation}
The no-penetration boundary condition remains unchanged. We must first account for the finite-slip boundary in our definition of $\psi_0$, which we do by considering flow in the absence of perioidic magnetic field. This is simply the following.

\begin{equation}
	\psi_0 = y  + \frac{e^{-y}-e^{y}}{e^{w/2}(1+\beta)+e^{-w/2}(1-\beta)}
\end{equation} 

While the equations for the free solution, Eq.~\eqref{app_eigenproblem} remain unchanged, the equations for the driven solution, Eq.~\eqref{app_driven} are modified by the redefinition of $\psi_0$.
	\begin{align} 
	(g^2m^2)(g^2m^2-1)f_m + \frac{\omega \tau g}{2}(f_{m+1} + f_{m-1})\\\nonumber = -\frac{\omega \tau g}{2(e^{w/2}(1+\beta)+e^{-w/2}(1-\beta))} (\delta_{m,1} + \delta_{m,-1})\\\nonumber
	(g^2m^2)(g^2m^2-1)h_m - \frac{\omega \tau g}{2}(h_{m+1} + h_{m-1})\\\nonumber = -\frac{\omega \tau g}{2(e^{w/2}(1+\beta)+e^{-w/2}(1-\beta))} (\delta_{m,1} + \delta_{m,-1})
\end{align}

Now, while the no-penetration boundary condition remains unchanged, we must solve the finite-slip boundary condition for $\delta \psi_d + \delta \psi_f$. The no-penetration and finite slip boundary conditions applied to the finite channel system are as follows.

\begin{align}
	\label{app_no_slip}
	&0=v_y|_{y=\pm W/2} \to\nonumber\\
	&m\bigg[f_m e^{\mp W/2} + h_m e^{\pm W/2} + \sum_a ( \alpha_aA_{m,a}^{(n)}e^{\mp\lambda_aW/2} \nonumber\\&+ \gamma_aC_{m,a}^{(n)}e^{\pm\lambda_aW/2} ) \bigg]+ \frac{\omega \tau}{2g(1+g^2)}(\delta_{m,1}-\delta_{m,-1}) = 0\nonumber\\
	\nonumber\\
	&v_x|_{y=\pm W/2}=\mp\beta \partial_y v_x|_{y=\pm W/2}\to\nonumber\\
	& \sum_a \lambda_a( \alpha_aA_{m,a}^{(n)}e^{\mp\lambda_aW/2}-\gamma_aC_{m,a}^{(n)}e^{\pm\lambda_aW/2}) \nonumber\\&+ f_m e^{\mp W/2} - h_m e^{\pm W/2} = \mp\beta \bigg[ f_m e^{\mp W/2} + h_m e^{\pm W/2} \nonumber\\&+\sum_a \lambda_a^2( \alpha_aA_{m,a}^{(n)}e^{\mp\lambda_aW/2}+\gamma_aC_{m,a}^{(n)}e^{\pm\lambda_aW/2})\bigg]  
\end{align}

\begin{figure}[ht!]
	\includegraphics[width=0.45\textwidth]{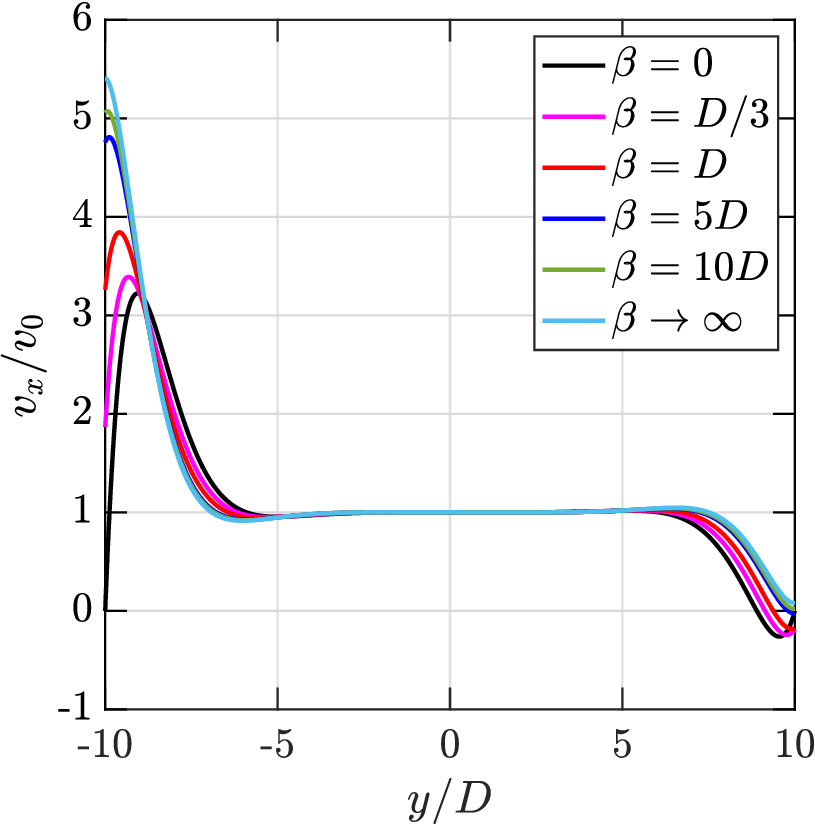}
	\caption{Longitudinal velocity profile at $x/D=0$ with $gD=0.88$ and $\omega\tau=4.8$ for a range of slip length $\beta$. Clearly the finite-slip cases interpolate between the no-slip and perfect-slip cases. As $\beta$ increases the strength of the anti-Poiseuille flow peak increases until it saturates at the perfect-slip level, while counterflow at the other boundary quickly disappears with finite $\beta$.
	}
	\label{vx_fs}
\end{figure}

It is instructive to compare the no-slip, finite-slip and perfect-slip cases. For monolayer graphene the slip length is dependent on temperature and has been found, as a fitting parameter, to be on the order of $0.5-0.1$ $\mu$ m for $T=300$K\cite{sulpizio_visualizing_2019} and $T=75K$\cite{ku_imaging_2020}. For the parameters we consider in this manuscript an appropriate choice is $\beta=D/3$. In Fig.~\ref{vx_fs} we present $v_x$ profiles for $\omega\tau=4.8$ for the different boundary conditions, with this $\omega\tau$ chosen to highlight anomalous flow behaviour. As the $\beta$ is increased the flow transitions from the no-slip case to the perfect-slip case, with the anti-Poiseuille flow peak increasing with increasing $\beta$ and counterflow at the opposite boundary quickly vanishing with small finite $\beta$. At realistic slip length $\beta=D/3$ one would expect strong periodic anti-Poiseuille flow as well as small finite counterflow.

\bibliography{Hydrodynamics}
\end{document}